\documentclass[11pt, preprint]{aastex}
\usepackage{amssymb,amsbsy}


\def\***#1{{\sc #1}}
\def\plan#1{\relax}
\def\Plan#1{\relax}
\def\PLAN#1{\relax}

\DeclareSymbolFont{euletters}{U}{eur}{m}{n}
\DeclareMathSymbol{\alpha}{\mathord}{euletters}{'013}

\def\etal{{\it et al.}}
\def\hi{{\rm HI}}
\def\be{\begin{equation}}
\def\ee{\end{equation}}
\def\b{\boldsymbol{\beta}}
\def\t{\boldsymbol{\theta}}
\def\dO{\Delta\Omega}
\def\M{M_{\rm HI}}
\def\DLA{{\rm  DLA}}
\def\mb{\bar{\mu}}
\def\la{\lower.5ex\hbox{$\; \buildrel < \over \sim \;$}}
\begin{document}

\title{Using Gravitational Lensing to study HI clouds at high redshift}

\author{Tarun Deep Saini,  }
\affil{Inter-University Center for Astronomy \& Astrophysics,
Pun\'e 411 007, India; saini@iucaa.ernet.in}

\author{Somnath Bharadwaj}
\affil{Department of Physics and Meteorology \& Center for Theoretical
Studies, I.I.T. Kharagpur, 721 302, India; somnath@phy.iitkgp.ernet.in}
\and
\author{Shiv K. Sethi} 
\affil{Harish Chandra Institute, Chhatnag Road, Jhusi, Allahabad
211 019, India; sethi@mri.ernet.in}

\begin{abstract}
We investigate the possibility of detecting HI emission from
gravitationally lensed HI clouds (akin to damped Lyman-$\alpha$
clouds) at high redshift by carrying out deep radio observations in
the fields of known cluster lenses. Such observations will be possible
with present radio telescopes only if the lens substantially magnifies
the flux of the HI emission. While at present this holds the only
possibility of detecting the HI emission from such clouds, it has the
disadvantage of being restricted to clouds that lie very close to the
caustics of the lens.  We find that observations at a detection
threshold of $50\,\,\mu {\rm Jy}$ at $320 \,\,\rm MHz$ (possible with
the GMRT) have a greater than $20\%$ probability of detecting an HI
cloud in the field of a cluster, provided the clouds have HI masses in
the range \mbox{$5\times 10^8 \,\,{\rm M}_{\odot} \le \M \le 2.5 \times
10^{10}\,\, {\rm M}_{\odot}$}. The probability of detecting a cloud
increases if they have larger HI masses, except in the cases where the
number of HI clouds in the cluster field becomes very small.  The
probability of a detection at $610\,\,{\rm MHz}$ and $233 \,\, {\rm
MHz}$ is comparable to that at $320 \,\, {\rm MHz}$, though a definitive
statement is difficult owing to uncertainties in the HI content at the
redshifts corresponding to these frequencies. Observations at a
detection threshold of $2\,\, \mu {\rm Jy}$ (possible in the future
with the SKA) are expected to detect a few HI clouds in the field of
every cluster provided the clouds have HI masses in the range $2
\times 10^7\,\, {\rm M}_{\odot} \le M_{\rm HI}
\le  10^9 \,\,{\rm  M}_{\odot}$. Even if such observations do not result
in the detection of HI clouds, they will be able to put useful
constraints on the HI content of the clouds.
\end{abstract}

{\it subject headings}: cosmology: gravitational
lensing; theory: galaxies; large-scale structure of the
universe; radio lines: HI

\section{Introduction}
Quasar spectra reveal a variety of absorption features superposed on
the continuum emission, indicating the existence of HI clouds with a
wide range of neutral hydrogen column densities distributed at
different redshift.  Among these the damped Lyman-$\alpha$ clouds
(DLA) which have HI column densities in the range $2 \times 10^{20}
\le N_{\hi} ({\rm cm^{-2}}) \le 10^{22}$ are the main repository of
neutral hydrogen at high redshift $(z \simeq 3)$. It has been
speculated that these systems are the progenitor of present day
spiral galaxies (for details see Wolfe 1995 and references therein).

 In addition to the column densities, absorption
studies have   determined the detailed velocity  structure of the
DLAs (Prochaska \& Wolfe 1998), and different  models 
have been proposed for DLAs in order to explain these observations. 
The DLAs have been modeled as thick rotating disks with rotation 
velocities in the range $200 \hbox{--}300 \,\, \rm km \, s^{-1}$
(Prochaska \& Wolfe 1998).  Another model  proposes that DLAs  are
made up of gaseous  protogalactic clumps undergoing in-fall 
(Haehnelt, Steinmetz \& Rauch 1997). 

Despite very detailed absorption studies, the exact nature of the DLAs
is still not fully understood. A part of the uncertainty arises from
the fact that absorption studies give information along a single line
of sight through a DLA, and this cannot be used to determine the total
HI content or the physical extent of these systems. In a few cases
there exist multiple lensed images of a quasar from which we can infer
that the angular size of the DLAs is $\simeq 3''$ (Smette 1995 and
reference therein). Observations of DLAs in HI absorption and
deep-imaging of the DLA fields in the optical band have also been used
to infer the size of DLAs (Lane {\it et al.} 2000; Fynbo {\it et al.}
2000; Fynbo {\it et al.} 1999; Moller {\it et al.} 1998; Warren \&
Moller 1996; Moller \& Warren 1993; Briggs {\it et al.}  1989; Briggs
1988), but such observations only give lower limits on the physical
size of DLAs.

In a few cases it has been possible to use deep optical imaging and
spectroscopy to  identify counterparts (in emission) of the systems
that produce the damped  Lyman-$\alpha$ lines. 
While most of these observations have been performed on low redshift
DLA fields (see e.g. Brun {\it et al.}  1997 and references therein)
there also exist several cases of positive detection of DLAs at high
redshift (Kulkarni {\it et al.} 2000;
Djorgovski {\it et al.}
1996; Fynbo {\it et al.} 2000; Fynbo {\it et al.} 1999;  Moller {\it
  et al.} 1998; Warren \& Moller 1996;  Moller \& Warren 1993).

Here we consider the possibility of detecting HI emission from
hydrogen clouds whose column densities are sufficiently high to
qualify as DLAs, if they were located along the line of sight to
quasars. This allows the properties of the HI clouds to be inferred
from the observed properties of DLAs.  These observations will also
enable the total HI content of the DLAs to be determined.
Unfortunately, the very small angular extent of the DLAs leads us to
expect HI flux from these clouds to be below the detection threshold
of existing radio telescopes.

Optical observations of background galaxies (typically $\la 1''$)
lensed by cluster of galaxies reveal arc-like images which could be as
large as $20''$ (see e.g. Schneider {\em et al. \/} 1992). Detailed
models have been constructed for these cluster lenses; and it is
possible to predict the magnification of the image of HI clouds if
they were located at different positions behind the cluster. The
lensing of HI clouds by clusters of galaxies holds the possibility of
magnifying them sufficiently, rendering the HI emission detectable by
presently available radio telescopes.  In this paper we investigate
the possibility of detecting the redshifted $1420 \,\, \rm MHz$ HI
emission from lensed HI clouds by carrying out deep radio observations
centered on known cluster lenses.

Our investigation is largely motivated by the fact that the Giant
Meterwave Radio Telescope (GMRT; G. Swarup {\it et al.\/} 1991) which
has several bands in the frequency range $150 \hbox{--} 1420 \, \,
{\rm MHz}$ has recently started functioning.  In this paper we
consider frequency bands of width $16 \,\,{\rm MHz}$ centered at $233,
\, 320$ and $ 620 \,\, {\rm MHz}$. These correspond to the redshifted
$1420 \,\, {\rm MHz}$ emission from HI at $z \simeq 1.3, \, 3.4$ and
$5.1$ respectively. Most of the discussion in this paper refers to
$320 \,\, {\rm MHz}$ which we have used as the fiducial frequency.  In
addition we use a spatially flat $\Omega_m=1$ model, and a Hubble
parameter $H_0=100 \,\, h
\, {\rm km \, s^{-1}}$. We shall use $h=0.5$ whenever a numerical
result is reported.

The absorption studies of DLAs have yielded a maximum HI column
density of $\la 10^{22}$ (Lanzetta, Wolfe \& Turnshek 1995).  It is
possible that clouds with higher column densities contain larger
amount of dust that obscures the background quasar, in which case
these clouds would not be detected in absorption (Fall \& Pei
1995). This possibility has been taken into account while analyzing
the possibility of detecting the HI emission by setting the maximum HI
column density of the clouds to a value which is a few times larger than
the maximum HI column densities observed in absorption.

In \S~2 we review the observed properties of HI clouds which are
relevant for our analysis and describe two simple models for the HI
cloud mass and flux distribution.  In \S~3 we briefly discuss a few
relevant features of gravitational lensing by a cluster of galaxies
and outline the method of our calculation.  We present our results in
\S 4 and in \S 5 we give our conclusions.

\section{Mass and flux distribution of HI clouds}
We begin this section with   a  brief discussion of  some of the
observational features of DLAs  which we use to determine the properties
of the distribution of HI clouds. 

Absorption studies directly give the velocity dispersion $\Delta V$ of
the HI along lines of sight through DLAs and these observations
indicate rotational velocities in the range $200 \hbox{--} 250 \,\, {\rm
km \,s^{-1}}$ (Prochaska \& Wolfe 1998). This along with the HI
column density can be used to calculate the specific intensity of the
redshifted $1420 \,\, \rm MHz$ emission from the HI along the line of
sight through a HI cloud. This gives (for derivation  see Appendix A):
\be
I_{\nu}= 40 \,\, \mu  {\rm Jy \,  arcsec^{-2}} (1+z)^{-3} 
\left( \frac{N_{\rm HI}}{10^{22} \,{\rm cm^{-2}}} \right)  \, 
\left(  \frac{200 \,\, {\rm km \, s^{-1}}}{\Delta V}  \right) 
 \,.
\label{eq:a1}
\ee
Integrating the specific intensity  over the angular extent of the HI cloud
gives the total  flux from the cloud: 
\be
F_{\nu} \propto
\int N_{\rm HI}(\b) \,\, d^2 \beta \,\,,
\label{eq:a2}
\ee
where $\b$ (in arc-seconds) refers to different angular positions on
the sky, and the integral is over the angular extent of the HI cloud.

In a situation where there is gravitational lensing a source which
would otherwise appear at $\b$ is seen at a different angular
position $\t$ (see e.g. Schneider {\it et al.} 1992), and the flux from the
lensed image of the HI cloud can be calculated by performing a similar
integration over the angular extent of the image as in Eq.~(\ref{eq:a2}).

Estimating the flux expected from a HI cloud (lensed or unlensed)
requires knowledge about its angular extent and the variation in the
column density across the HI cloud. As these facts are not available,
we have adopted two simple models for our calculations in this paper. 
We describe these models below.

\subsection{Uniform disk (UD)}
In this model the HI clouds are assumed to be face on circular  disks, all
of the same physical radius $r_{\rm DLA}$ and with uniform column density
across the disk.
We consider different values of $r_{\rm DLA}$ in the range $10\,\, {\rm
kpc}$ to  $30  \,\, {\rm kpc}$. The solid angle subtended by a HI cloud is
$\dO_{\rm DLA}=\pi r^2_{\rm DLA}/d^2_s(z)$ where $d_s(z)$ is the 
angular diameter distance to the HI cloud (see e.g. Peebles 1993).
In this model the  flux from an unlensed HI cloud is 
$F_{\nu}=I_{\nu} \dO_{\rm DLA}$  and we have, for $\nu = 320 \,\, {\rm
MHz}$, 
\be
F_{\nu}= 12 \,\, \mu {\rm Jy} \, h^2 \left(\frac{N_{\rm HI}}{10^{22} \,
{\rm cm^{-2}}} 
\right) \left(\frac{200 \, {\rm km \, s^{-1}}}{\Delta V}\right)
\left( \frac{r_{\rm DLA}}{10 \,  {\rm kpc}} \right)^2.
\label{eq:a7}
\ee

The distribution of the column densities of the HI clouds is assumed  to be
a power law $F(N_{\rm HI})= B N^{-\alpha}_{\rm HI}$ with values in the
range  $2 \times 10^{20} \, {\rm cm^{-2}}< N_{\rm HI} < 10^{22} \,
{\rm   cm^{-2}}$, where  $F(N_{\rm HI})\, d  N_{\rm HI}$ gives  the
number of HI clouds  per unit volume with column density in the range $d 
N_{\rm HI}$.    Observations  indicate that the column density
distribution changes with redshift implying that both 
$\alpha$ and  $B$ change with $z$. The 
DLA  population shows a marked  increase in the HI column  densities
at higher redshifts (Lanzetta {\it et al.} 1995) and $75 \%$
of the DLAs in the redshift range $3 \le z \le  3.5$ have column
densities  $N_{\rm HI} \ge 10^{21} {\rm cm^{-2}}$. This is consistent
with $\alpha=0.5$. In our work we consider values of $\alpha$ in the range
$0.5$ to $1.0$. 

In this model the HI mass of a cloud is given by 
\begin{equation}
\M = 2.5  \times 10^{10} \, {\rm M_{\odot}} \left ({r_{\DLA}
    \over 10 \, {\rm kpc}} \right )^2 \left ( {N_{\rm HI} \over
    10^{22} \, {\rm cm^{-2}}} \right ).
\label{eq:a8}
\end{equation}
This relation permits us to interchangeably use either a column
density distribution function or a HI mass distribution function
$f(\M)=A \M^{-\alpha}$ where $f(\M) d \M$ is the number of HI clouds
per unit volume with HI mass in the range $d \M$. The mass range
corresponding to the column density range $2 \times 10^{20} \, {\rm
cm^{-2}}< N_{\rm HI} < 10^{22} \, {\rm cm^{-2}}$ depends on the value
of $r_{\DLA}$, and for $r_{\DLA}= 10 \,\, {\rm kpc}$ we have $\M[{\rm
min}] = 5 \times 10^8 \,\,{\rm M}_\odot$ and $\M[{\rm max}] = 2.5
\times 10^{10} \,\, \rm M_\odot$.
At any redshift $z$ the HI mass distribution function can be used to
calculate the HI density
\be
\int_{\M[{\rm min}]}^{\M[{\rm max}]} \M \, f(\M) \, d  \M = \rho_c \,
\Omega_{\rm HI}  \,\,,
\label{eq:a5}
\ee
where $\rho_c$ is the critical density at redshift $z$ and
$\Omega_{\rm HI}$ is the contribution to the density parameter from
the HI in DLAs at that redshift.  Absorption studies (Lanzetta
{\it et al.} 1995) have shown that in a flat cosmological model, in
the redshift range $z \la 3.5$, the total mass density contributed by
the HI in DLAs can be fitted by the function
\be
\Omega_{\rm HI}(z) = 0.19 \times 10^{-3}  h^{-1}\exp (0.83 z)\,\,.
\label{eq:a4}
\ee
We use this to fix the normalization constant $A$ for the HI mass
distribution  function.

\subsection{Exponential disk (ED)}
Recent studies of the velocity structure of the DLAs show that the HI
density across the cloud is consistent with a thick, rotating disk of
exponential profile (Prochaska \& Wolfe 1998)
\begin{equation}
n(R,Z) = n_0 \exp\left(-\frac{R}{R_{\rm d}} - 
\frac{|Z|}{Z_{\rm d}} \right)\,\,,
\label{eq:a9pp}
\end{equation}
where $n(R,Z)$ is the number density of HI atoms in the disk, $R$ is
the radial coordinate in the plane of the disk and $Z$ is the
coordinate along the thickness of the disk.  Observations suggest that
disks have $R_{\rm d}/Z_{\rm d} \simeq 3$, and are rotating with $v_c
\simeq {200\hbox{--}250} \,\, \rm km \, s^{-1}$.
Excluding the effect of rotation, the gas in the disks has negligible
velocity dispersion. Absorption studies do not fix $R_{\rm d}$ and
they are only mildly dependent on the central column density $N_{\rm
HI}(0) = 2\, Z_{\rm d} n_0$. In our analysis we assume that all the
disks are observed face-on. As a consequence the column density
distribution in each HI cloud is circularly symmetric around the
center and falls off exponentially in the radial direction.  For $\nu
= 320 \,\,\rm MHz$, the flux from a HI cloud is
\be
F_{\nu} = 24 \,\, \mu  {\rm Jy} \,  h^2 \left(\frac{N_{\rm HI}(0)}{
10^{22} \, 
{\rm cm^{-2}}} 
\right) \left(\frac{200 \,\,{\rm km \,s^{-1}}}{\Delta V} \right)
\left( \frac{R_{\rm d}}{10 \,\,{\rm kpc}} \right)^2 \,\,,
\label{eq:a9}
\ee
and the  HI  mass is
\begin{equation}
\M = 6 \times 10^{10}\,\, {\rm M}_\odot \left
({N_{\rm HI}(0) \over 10^{22} \,\, {\rm cm^{-2}} } \right ) \left
({R_{\rm d} \over 10 \,\, {\rm kpc} }\right)^2 \,\,.
\label{eq:a10}
\end{equation}
We choose the central column density, $N_{\rm HI}(0)$ to have a fixed
value in the range $5 \times 10^{21} \, \rm cm^{-2}$ to $2 \times
10^{22} \, \rm cm^{-2}$ and allow $R_{\rm d}$ to vary from $2\, \, \rm
kpc$ to $10 \, \, \rm kpc$.  For a fixed value of $N_{\rm
HI}(0)=10^{22} \,\, \rm cm^{-2}$ this corresponds to a mass range
$\M[{\rm min}] = 2\times 10^9 \,\, {\rm M_\odot}$ and $\M[{\rm max}] = 5
\times 10^{10}\,\, {\rm M_\odot}$. 
The treatment of the mass distribution function for the ED model has
been carried out in exactly the same way as for the UD model, and we
consider values in the range $0.5$ to $1.0$ for the slope $\alpha$.

\subsection{The flux distribution of HI clouds}
For both the UD and the ED model the HI mass distribution function can
be converted to a flux distribution function using either
Eqs.~(\ref{eq:a7}) and~(\ref{eq:a8}) or Eqs.~(\ref{eq:a9})
and~(\ref{eq:a10}).  This requires assuming a value for $\Delta V$
which we take to be $200 \,\, \rm km \, s^{-1}$.  The GMRT channel at
$320 \,\, {\rm MHz }$ $(z=3.4)$ has a bandwidth of $16 \,\, {\rm
MHz}$. This corresponds to a redshift range of $3.33$ to $3.55$. We
use the flux distribution function to calculate the number of clouds
per unit flux per $\rm arcsec^2$ within the redshift range
covered by the GMRT bandwidth.

For the UD model where the value of $r_{\DLA}$ is fixed this
gives us  

\begin{eqnarray}
\nonumber
&&{d^2 n \over d F_{\nu} d\Omega} \simeq 1.2 \times 10^{-3} {\rm \mu Jy^{-1} \, arcsec^{-2}}   \times\\
&& \,(2-\alpha) \left ({r_{\DLA} \over 10 \,
{\rm kpc}} \right)^{-4 + 2 \alpha} \left ( { F_{\nu} \over 1\, \rm \mu Jy}
\right)^{-\alpha} \,\,.
\label{eq:a11}
\end{eqnarray}
For this model the different fluxes correspond to different values of
the column density. For the ED model where the central column density
$N_{\rm HI}(0)$ is fixed,   
the number of clouds per unit flux per $\rm arcsec^2$
is 
\begin{eqnarray}
\nonumber
&&{d^2 n \over dF_{\nu} d\Omega} \simeq 4.25  \times 10^{-4} {\rm \mu Jy^{-1} \, arcsec^{-2}}  \times \\  
&&\,(2-\alpha) \left ({N_{\rm HI}(0) \over 10^{22}
\, {\rm cm^{-2}}} \right)^{-2 +  \alpha} \left ( { F \over 1\, \rm \mu
Jy} \right)^{-\alpha} \,\,.
\label{eq:a12} 
\end{eqnarray}
(In Eqs.~(\ref{eq:a11}) and~(\ref{eq:a12}) the normalization has a weak
dependence on the parameter $\alpha$, which is not shown for brevity)
In this model the variation of flux corresponds to different values of
the radial scale length $R_{\rm d}$. For $\nu = 320 \,\, {\rm MHz }$,
the total number of clouds expected above a given flux in a $1'\times
1'$ region of the sky and within the GMRT bandwidth is shown in
Figure~\ref{fig:ud1} for the UD and ED models.

\section{Gravitational Lensing by clusters}
Here we give a brief summary of the salient features of gravitational
lensing relevant to this work. A more thorough exposition can be found
in Schneider {\em et al. \/} (1992).

An object which is gravitationally lensed appears at a position which
is different from its unlensed position. This information is
encoded in the lens equation which relates the unperturbed position of
the source to its perturbed position.  If the angular coordinates of
the source are $\boldsymbol \beta$ and the angular coordinates of the image
are $\boldsymbol \theta$, then the lens equation is given by
\begin{equation}
\boldsymbol{\beta } = \boldsymbol {\theta } - \boldsymbol{{\nabla}} \psi 
(\boldsymbol{ \theta }) \,\,.
 \label{eq:lens}
\end{equation}
The dimensionless relativistic lens potential $\psi$ satisfies the
two-dimensional Poisson equation $ \nabla ^2 \psi ( \boldsymbol{ \theta }) = 2
\kappa( \boldsymbol{\theta })$, where the  convergence $\kappa( \boldsymbol
{ \theta}) = \Sigma ( \boldsymbol { \theta })/ \Sigma _{\rm cr}$ and
$\Sigma (\boldsymbol{\theta})$ is the two-dimensional surface mass
density of the lens, $\Sigma _{\rm cr}= (c^2/4 \pi G)(d_s/d_ld_{ls})$
being the critical density. The quantities $d_s$, $d_l$, and $d_{ls}$
are the angular-diameter distances from the observer to source, the
observer to lens, and the lens to source respectively.  Gravitational
lensing does not effect the specific intensity along a light ray. The
flux received by an observer is proportional to the solid angle
subtended by the image at the observer; and since the solid angle of
the image after lensing is, in general, different from that of the
source, an observer can receive more (or less) flux in the lensed case
than in the unlensed case.

The shape and size of the image are related to those of the source by
the transformation matrix $ M^{-1}_{ij} = \partial \beta_i/ \partial
\theta_j$. The infinitesimal area $d^2 \theta$ in the image plane
is mapped to an area $d^2 \beta=
\mu^{-1}(\boldsymbol{\theta}) d^2 \theta$ in the source plane, where $
\mu = {\rm det}[M_{ij}]$ is the magnification.  The regions (curves)
in the source plane where $\det[\partial \beta_i/
\partial \theta_j] = 0$ are called caustics. Magnification is
infinite for point sources which are placed on the caustics. For
finite sources the amplification remains finite, though the caustics still
remain the points in the source plane where the magnification is large.

For this paper we model clusters of galaxies as single component
Pseudo Isothermal Elliptical Mass Distribution (PIEMD) after Kassiola
\& Kovner (1993).  These are characterized by a core radius $R_{\rm c}$,
one-dimensional velocity dispersion $\sigma$, and ellipticity
parameter $\epsilon$. In addition to these parameters, the lens
redshift $z_l$ and the source redshift $z_s$ completely specify the
dimensionless potential $\psi$. Real clusters can have
substantial sub-structure and are best modeled as multi-component
PIEMDs, but for simplicity we consider only single component lenses.

For our case the source redshift, $z_s =\nu_e/\nu_o  -1$ with $\nu_e =
1420 \, \rm MHz$, and $\nu_o = \{ 233,\, 320, \, 620\} \,  \rm MHz$. 
For our study we consider the following range of parameters: $1200 \,\,
{\rm km \,s^{-1}} \le \sigma \le 1800 \,\, {\rm km \,s^{-1}}$; $40
\,\, {\rm kpc} \le R_{\rm c} \le 60 \,\, {\rm kpc}$; $0 \le \epsilon \le 0.3$,
and $0.1 \le z_l \le 0.3$. This choice of parameters is motivated by
the observed properties of known clusters and their abundances
(e.g. Mazure \etal,  1996).

\subsection{Lensing of HI clouds}
The flux from a lensed HI cloud  can be expressed in
terms of the source coordinates $\b$  and the magnification
$\mu(\b)$ as:
\begin{eqnarray}
F_{\nu}&=& 40 \,\,\mu  {\rm Jy\,  arcsec^{-2}}  (1+z)^{-3} 
\left(  \frac{200 \,\, {\rm km \, s^{-1}}}{\Delta V}  \right) \nonumber \\
{}&& \times \int \left( \frac{N_{\rm HI}(\b)}{10^{22}\,\, {\rm cm^{-2}}}
\right) \, \, \mu(\b) d^2 \beta \,\,.
\label{eq:b1}
\end{eqnarray}
This can be expressed in terms of the flux of the unlensed source
\be
F_{\nu}[{\rm Lensed}] =F_{\nu}[{\rm Source}]~\times~\bar{\mu} \,\,,
\ee
where $\bar{\mu}$ is the  magnification averaged over the angular
extent of the source   
\begin{equation}
\bar{\mu}(\b,R)  = \int d^2 \beta^{'} \, \mu(\b -\b') \, W(\b', R)\,\,. 
\label{eq:b2}
\end{equation}
The normalized window function $W(\b,R)$ represents the radial
profile of the HI cloud and for the UD and ED models we use

\begin{eqnarray}
W(\b,r_{\rm DLA})& =&
\frac{d_s^2}{\pi r_{\rm DLA}^2}  \Theta \left (|\b|  - \frac{r_{\rm  DLA}}{d_s}\right )
\qquad ~~    ({\rm UD})  \nonumber \\
W(\b, R_d)&=& \frac{d_s^2}{ 2 \pi R_{\rm d}^2} 
\exp \left(- \frac{|\b|}{R_d/d_s}\right) 
\qquad ({\rm ED})\,\,,
\end{eqnarray}
where $\Theta$ is the Heaviside step function. In our calculations we
have evaluated $\bar{\mu}(\b,R)$ by first calculating $\mu(\b)$ on a
finely-spaced grid in the source plane. This is then convolved with
the window function using the Fast Fourier Transform algorithm (Press
{\it et al. \/} 1992) to obtain $\bar{\mu}(\b,R)$. We have tested the
convergence of this procedure with respect to variation of the grid
size. In situations where there are multiple images we have added up
the magnification of all the magnified images; because the typical
image separation is less than the beam width of the GMRT and the
observations which we are discussing here will not be able to resolve
the multiple images.

We use the procedure discussed above to calculate $\bar{\mu}(\b,R)$ on
a grid. This is used to calculate $A( > \bar{\mu}_0)$
($\bar{\mu}_0$ is any fiducial average magnification) which is the
area (in ${\rm arc\hbox{-}second^2}$) in the source plane for which the
condition  $\bar{\mu}(\b,R) >  \bar{\mu}_0 $ is satisfied.  For
observations with a detection threshold of $F_{\nu}[{\rm min}]$,
a HI cloud with unlensed flux $F_{\nu}$ will have a lensed flux greater
than $F_{\nu}[{\rm min}]$ for any position within the area
$A(>F_{\nu}[{\rm min}]/F_{\nu})$. Summing up the contributions from
HI clouds with all possible values of the  unlensed flux gives us 
 $ {\cal N}({>} F_{\nu}[{\rm min}])$ the total number of HI clouds that
can be  detected in the field of a cluster
\begin{equation}
{\cal N}({>} F_{\nu}[{\rm min}]) = \int d F_{\nu} \,
{d^2 N  \over d\Omega d F_{\nu}} \,
 A \left( >F_{\nu}[{\rm min}]/F_{\nu} \right)   \,.
\end{equation}  

It should be pointed out that throughout the analysis we have assumed
all the HI clouds to be oriented face-on. In reality the HI clouds
will occur at
all possible angles. In this case
the mean magnification $\bar{\mu}(\b,R)$ will differ
with the orientation of the cloud. However, 
the projected area is maximum when the cloud is face-on
and the mean magnification will be lower relative to the situation
where the cloud is located at the same position with an edge-on
orientation. This means that our analysis gives an underestimate of
the magnification and we may expect larger magnifications when all
possible orientations are taken into account.

\section{Results}
We first discuss our results for observations centered at $320 \,\, {\rm
MHz}$. The results for $620$ and $233 \, {\rm MHz}$ are summarized at
the end of this section.   

\begin{figure}
\includegraphics[angle=0, width=\textwidth]{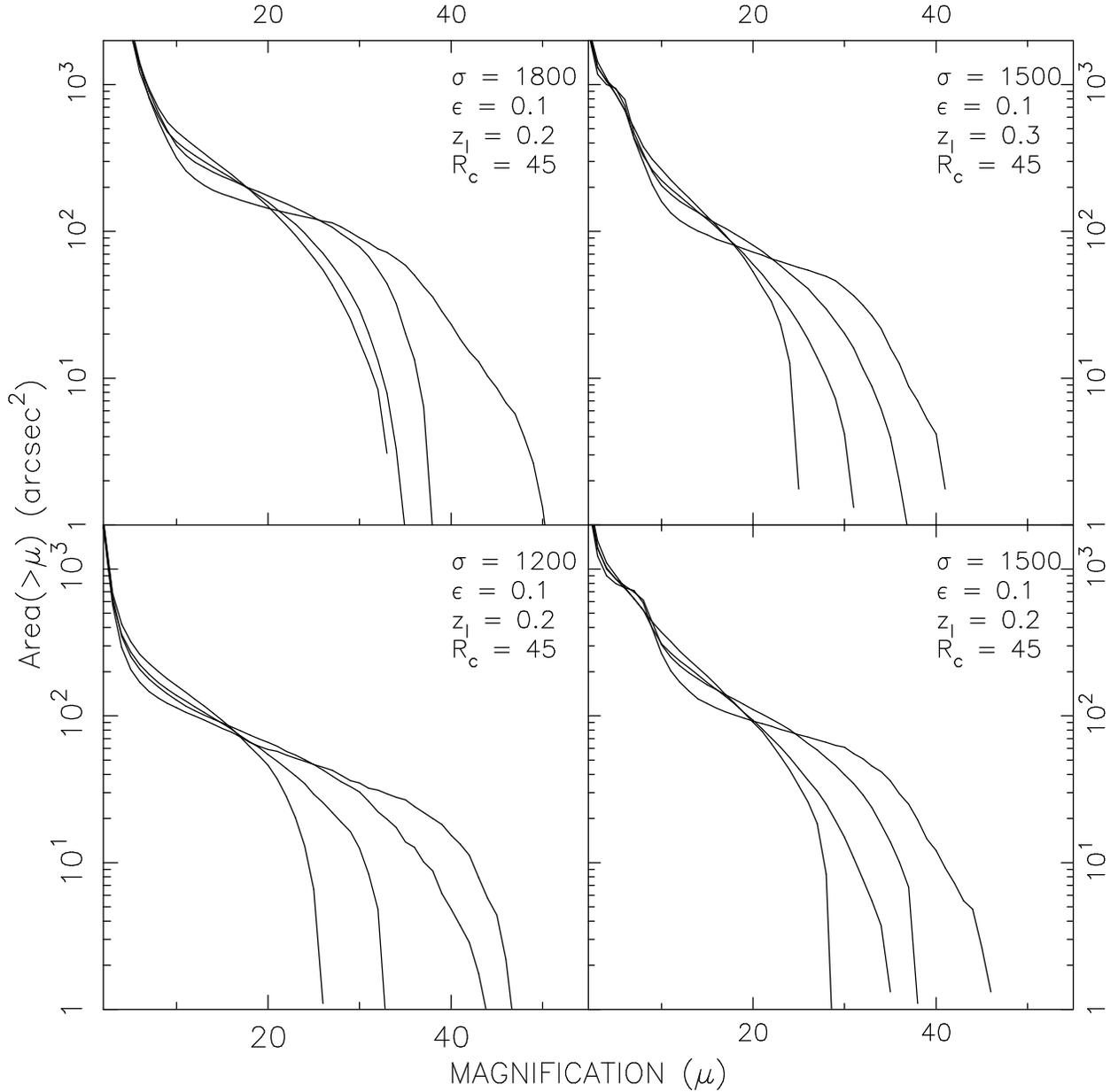}
\caption {The area in which the averaged amplification $\bar{\mu}$ exceeds a
given value $\mu$ is shown for several cluster parameters and clouds
radii in the the UD   model. In each panel, the clouds radius  $
r_{\rm  DLA} = \{10, 20,30, 40 \} \,\, \rm kpc$ corresponds to curves of
  decreasing maximum  amplification. }
\label{fig:ud2}
\end{figure}

\begin{figure}
\includegraphics[angle=0, width=\textwidth]{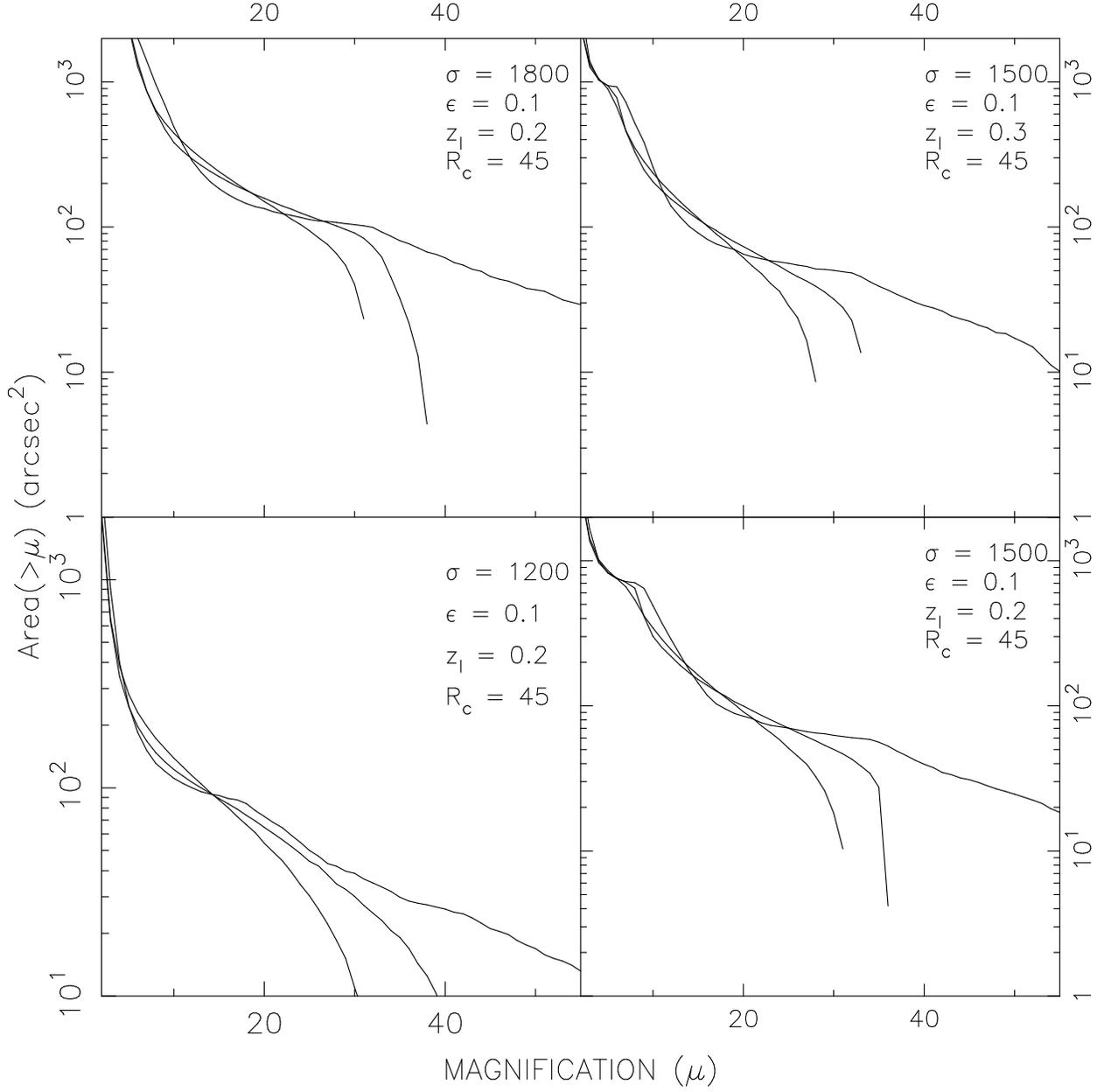}
\caption {The area in which the averaged amplification $\bar{\mu}$ exceeds a
given value $\mu$ is shown for several cluster parameters and radius
parameter $R_d$ in the the ED model. In each panel,
the radius 
parameter $R_d = \{2, 8, 14 \}  \,\, \rm kpc$, corresponding to curves
of   decreasing maximum  amplification. }
\label{fig:ed2}
\end{figure}

We have calculated $A({>}\mb)$ and ${\cal N}({>} F_{\nu}[{\rm min}])$
for several cluster parameters with the disk parameters varying over
the allowed range discussed earlier for the UD and ED models.  The
effect of varying the cluster parameters is to change $A({>}\mb)$
which in turn affects the number of detected HI clouds.  The regions
of high magnification ($\mu>10$) are restricted along the caustics and
all the contribution to $A({>}\mb)$ is from a narrow strip along the
caustics. Although the length of the caustic curve increases as the
ellipticity $\epsilon$ is increased, this is accompanied by a fall in
the values of the magnification.  The value $\epsilon=0.1$ is a good
compromise between these two effects and the results are presented for
this value.  The $\epsilon$ dependence of the results is rather weak
in the range $0.1 \le \epsilon \le 0.3$. The values of $A({>}\mb)$ and
${\cal N}({>} F_{\nu}[{\rm min}])$ decrease by around $10 \%$ as the
value of $\epsilon$ changes from $0.1$ to $0.3$.  Amongst the cluster
parameters, changing the velocity dispersion $\sigma$ has the
strongest effect on $A({>}\mb)$ and this is shown in
Figures~\ref{fig:ud2} and \ref{fig:ed2}.  Varying the cluster redshift
from $0.1$ to $0.5$ causes $A({>}\mb)$ to decrease by nearly $30\%$.

\begin{figure}
\includegraphics[angle=0, width=\textwidth]{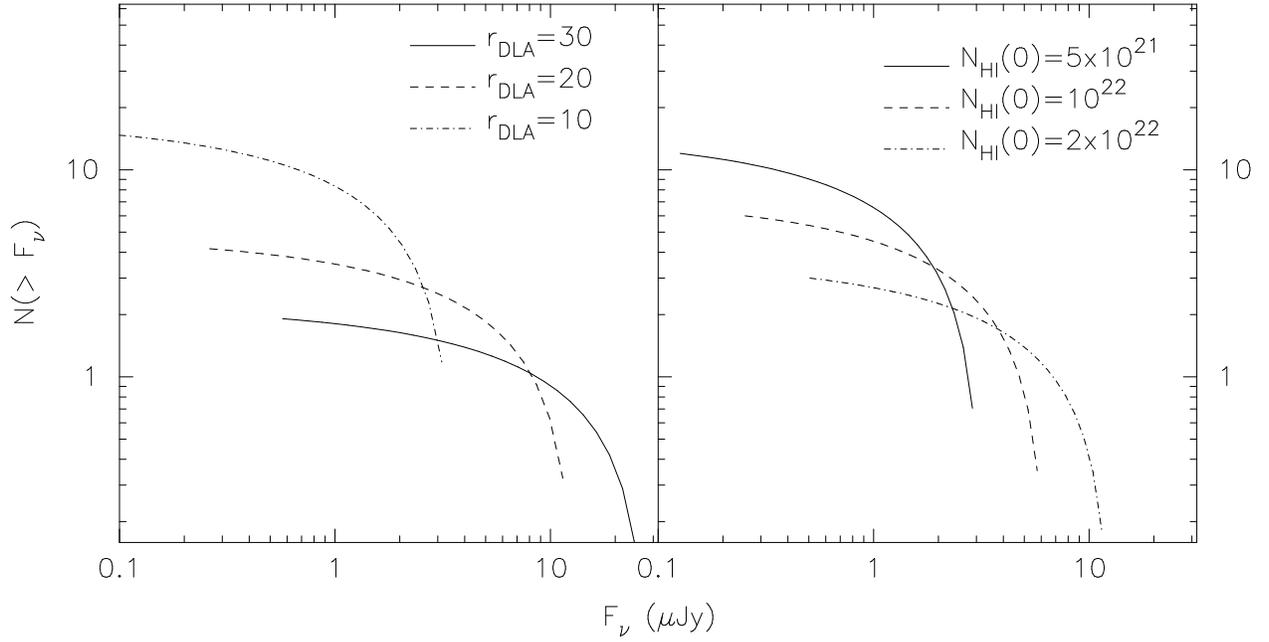}
\caption {
This shows the  number of HI clouds with flux greater than $F_{\nu}$
in a $1' \times 1'$ patch of the sky for a bandwidth of $16\,\, {\rm MHz}$
centred at $320 \,\, {\rm MHz}$, and a  mass spectral index $\alpha
= 0.5$. {\it Left Panel} is  the UD model
for three values of  $r_{\rm DLA}$ (in kpc). {\it Right Panel} is the ED model 
for three values of $N_{\rm HI}(0)$.}
\label{fig:ud1}
\end{figure}

\begin{figure}
\includegraphics[angle=0, width=0.9\textwidth]{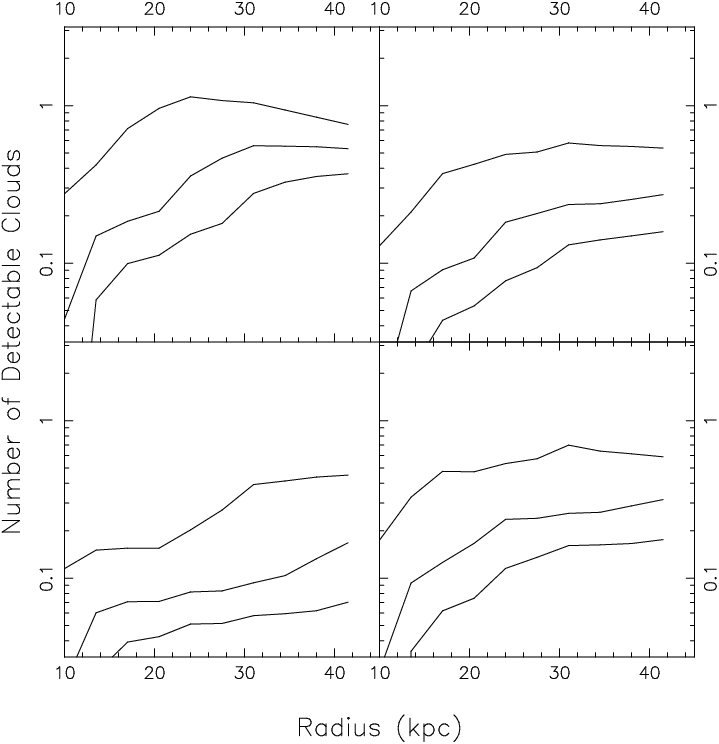}
\caption{This shows the number of lensed HI clouds expected to be
detected  in the field of a cluster as a function of the 
cloud radius in the UD model with $\alpha=0.5$,  for the 4 cluster
models shown in 
Figure~\ref{fig:ud2}. The three curves in each panel correspond to 
detection  thresholds $ F_{\nu}[{\rm min}]= \{ 50 , 100, 150 \} \,\, \rm
\mu Jy$, with the number of expected detections decreasing for higher 
flux thresholds.}
\label{fig:ud3}
\end{figure}

\begin{figure}
\includegraphics[angle=0, width=0.9\textwidth]{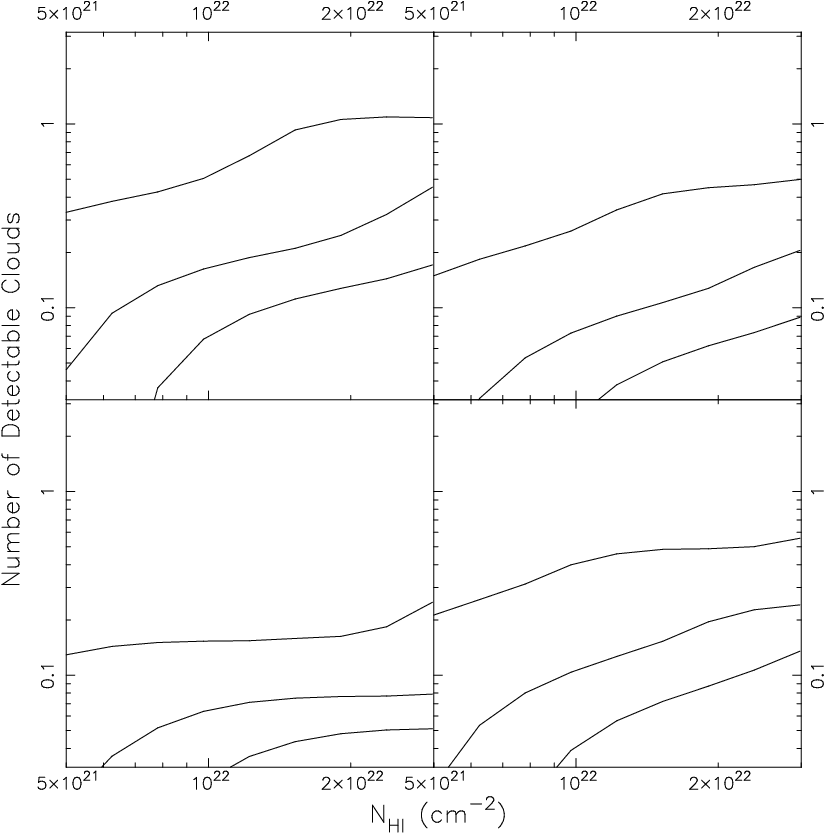}
\caption {This shows the number of lensed HI clouds expected to be
detected in the field of a cluster as a function of
the  central column density $N_{\rm HI}(0)$ in the ED model with 
 $\alpha = 0.5$. for the  
4 cluster  models shown in Figure~\ref{fig:ed2}. The three curves in
each panel correspond to  detection thresholds $F_{\nu}[{\rm min}]= 
\{ 50 , 100, 150 \} \,\, \rm \mu  Jy$ with the number of expected
detections decreasing for higher flux thresholds. }
\label{fig:ed3}
\end{figure}

Varying the  radius of the HI clouds has a few distinct
effects. First, an increase in the size of the HI clouds
decreases  the average magnification $\bar{\mu}(\b,R)$.
The area $A({>}\mb)$ also decreases as the radius of
the HI cloud is increased and this is clearly seen in Figures~\ref{fig:ud2} 
and~\ref{fig:ed2}. The second effect is that a larger
cloud radius gives a larger value of the unlensed flux (Eqs.~( \ref{eq:a7})
and~(\ref{eq:a9})). The third effect is that as the radius of the HI clouds
is increased, the number of clouds in the field decreases. 
This is seen in Figure~\ref{fig:ud1} for the UD model.
All these effects combine to determine how the  number of detected HI
clouds depends
on the  radius of the  clouds.  
As is seen  in Figure~\ref{fig:ud3} for the UD model, an increase in
the cloud radius generally  increases the number of detected
HI clouds.  Only for large values of $r_{\DLA}$ do
we find a decrease in the number of HI clouds causing a decrement in the
number of detections. 
The results for the ED  model shown in Figure~(\ref{fig:ed3})  are
qualitatively similar to those for the  UD model. 
In this case the parameter that controls the number of clouds 
in a field is $N_{\rm HI}(0)$, and  
the number of clouds in a field falls as  $N_{\rm HI}(0)$ is increased
(Figure~\ref{fig:ud1}).
As in the UD case, this  generally  leads to an 
increase in the number of detected clouds 
because a fall in the number of clouds is often 
over-compensated by an increase in the unlensed flux.
An important parameter in both the cloud models is the spectral
index, $\alpha$ (Eqs.~(\ref{eq:a11}) and~(\ref{eq:a12})). The number
of detected clouds decreases by   $\la 25\%$ as $\alpha$ is increased
from $0.5$ to $1$.

For a detection threshold of $50 \,\, \mu {\rm Jy}$, for observations
centered on a cluster with $\sigma=1200 \,\, \rm km \, s^{-1}$, the
expected number of detected clouds is in the range $ 0.1$ to $0.5$ for
$r_{\DLA} \ge 10\,\, {\rm kpc}$ in the UD model and for $N_{\hi}(0)
\ge 5 \times 10^{21} {\rm \, cm^{-2}}$ in the ED model.  For both the
UD and the ED models the lower limit for detection approximately
corresponds to the mass range $5
\times 10^8 \,\,{\rm M}_{\odot} \le \M \le 2.5 \times 10^{10} \,\,{\rm M}_{\odot}$.

The number of detections will go up if the observations are
centered on a cluster with a higher velocity dispersion, since ${\cal N}(>
50 \,\,\mu {\rm Jy})$ goes up by a  factor of $2$ to $3$ if  $\sigma=1800 \,\,
{\rm km\, s^{-1}}$  instead of  $\sigma=1200 \,\, {\rm km\,s^{-1}}$
(Figures~\ref{fig:ud3} and~\ref{fig:ed3}).
In addition it will be possible to detect the HI clouds even if they have lower
HI masses.

\begin{figure}
\includegraphics[angle=0, width=0.9\textwidth]{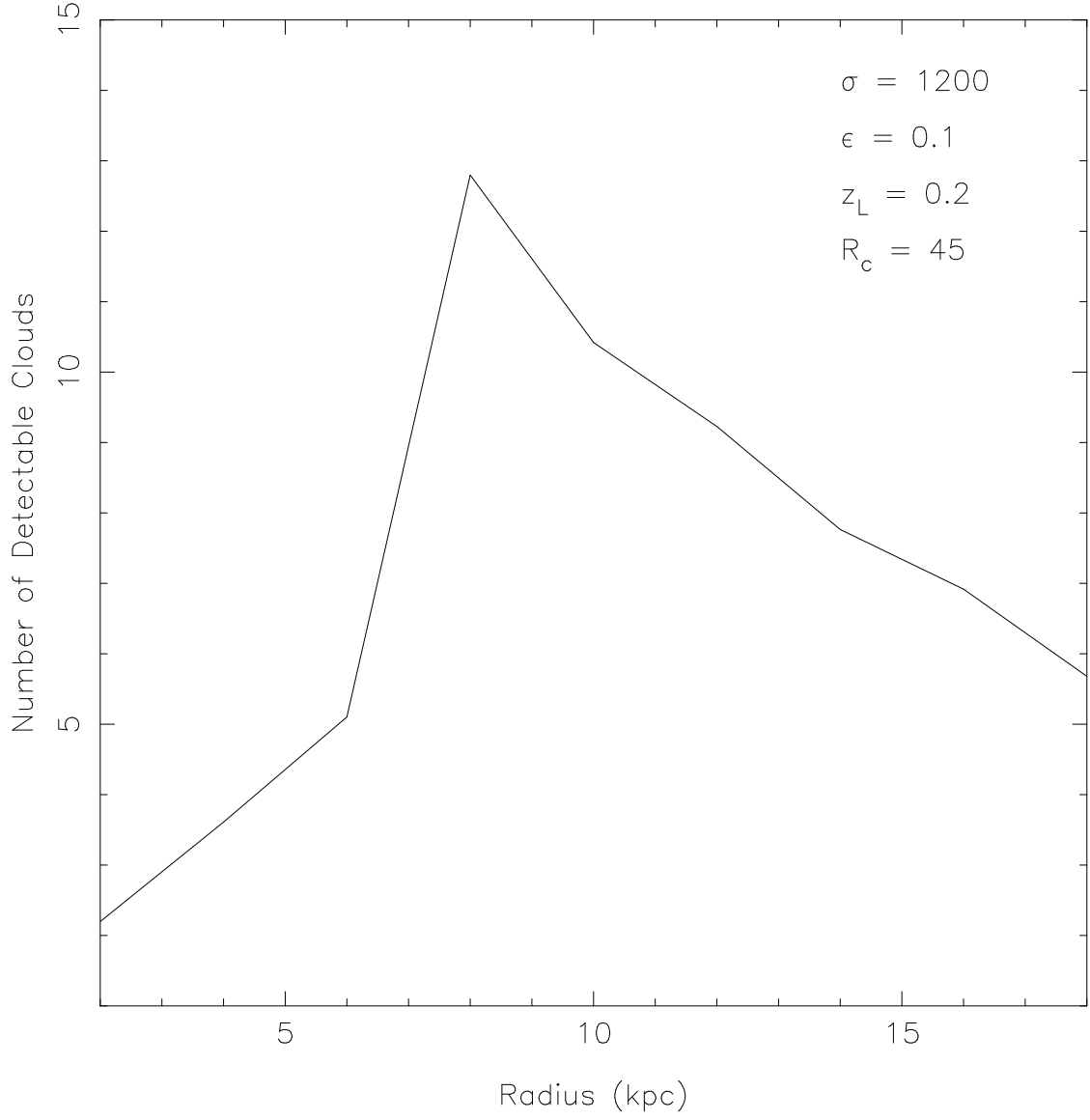}
\caption{This shows the number of lensed HI clouds expected to be
detected  in the field of a cluster as a function 
of the cloud radius, $r_{\rm DLA}$, in the UD model with $\alpha=0.5$ 
for observations with 
$ F_{\nu}[{\rm min}] = 2 \,\, \rm \mu Jy$ and a bandwidth of $16 \,\,{\rm
MHz}$ centred at $320 \,\, {\rm MHz}$. The cluster parameters are shown in
the figure. }
\label{fig:ska}
\end{figure}

Future radio telescopes  will reach sensitivities of around $2 \, \rm
\mu Jy$.  
Eqs.~(\ref{eq:a7}) and~(\ref{eq:a9}) show that it may then be
possible to detect  the HI clouds without the aid of gravitational lensing. 
In the UD model if $r_{\DLA} \ge 20 \,\, {\rm kpc}$ then a substantial
fraction  of the HI clouds will be detected without any  gravitational
lensing. While a small fraction of the  HI clouds may be detected without
gravitational  lensing if  $10\,\, {\rm kpc} < r_{\DLA}  < 20 \,\,{\rm
kpc}$, it will not be possible to observe the clouds without the aid of   
gravitational lensing  if  $r_{\DLA} \le \, {10 \,\,\rm kpc}$.
Figure~\ref{fig:ska} shows that observations with a detection
threshold of $2 \, \rm \mu Jy$ centered on a  cluster with $\sigma =
1200 \,\,{\rm km \, s^{-1}}$ are expected to have a few detections even
if the cloud radius is as small as $r_{\DLA}=2 \,\, {\rm kpc}$.  This
corresponds to a mass range $2 \times 10^7\,\, {\rm M}_{\odot} \le M_{\rm HI}
\le  10^9 \,\,{\rm M}_{\odot}$.

The discussion until now has been restricted to $320 \,\, \rm MHz$. 
The main differences which arise at other
frequencies are: 
(a) the flux from HI clouds  changes (Eq.~(\ref{eq:a2})), (b) the total number
of clouds is different owing to changes in  $\Omega_{\rm HI}$
(Eq.~(\ref{eq:a4})) and $\Delta z$ and,  (c) the  magnification of
sources changes owing to change in the source redshift (\S 3).

\begin{figure}
\includegraphics[angle=0, width=0.9\textwidth]{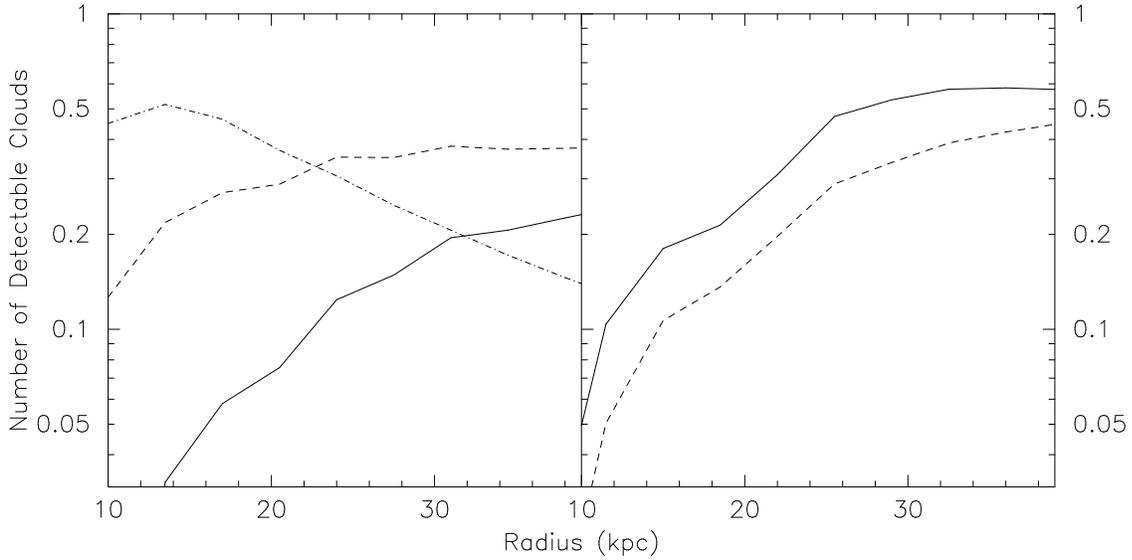}    
\caption{This shows the number of lensed HI clouds expected to be
detected in the field of a cluster with  $\sigma = 1800 \,\, \rm km \, 
 s^{-1}$, $\epsilon = 0.1$, $z_l = 0.2$ and $R_c = 45 \,\, \rm kpc$ in
the UD model. {\it Left Panel} is at $610 \,\, {\rm MHz}$. The solid
curve is for $N_{\rm HI}[max] = 10^{21} \,\, {\rm cm}^{-2}$, 
$\alpha = 2.3$ and $\Omega_{\rm HI}$ given by Eq.~(\ref{eq:a4}).
The dashed (dot-dashed) curve uses  $\alpha = 2.3$ ($\alpha = 0.5$)
with $\Omega_{\rm HI}$  twice the prediction of eq. (\ref{eq:a4}) and
$N_{\rm HI}[max] = 10^{22} \,\, {\rm cm}^{-2}$
{\it Right Panel} is at $233 \,\,{\rm  MHz}$ with 
$\alpha = 0.5$ (solid line) and $\alpha = 1.5$  (dashed line) }
\label{fig:othchan}
\end{figure}

At $z\simeq 1.3$ $(610 \,\, {\rm MHz})$ observations indicate $\alpha
\simeq 2.5$ and $N_{\rm HI}[\rm max] \sim 10^{21} \, \rm cm^{-2}$
(Lanzetta {\it et al.} 1995). Also, $\Omega_{\rm HI}$ falls
considerably from its value at $z \simeq 3.4$
However, owing  to obscuration of quasar light by the  dust present
in HI clouds (Fall \& Pei 1995)  these values may not be 
representative  of the population of HI clouds and it is
possible that the observed  $\Omega_{\rm HI}$  is
underestimated by a factor  of 3 (Fall \& Pei 1995) at this
redshift . Given the uncertainty in the  HI content we have used 
several values of spectral indices and $\Omega_{\rm HI}$.
Figure~(\ref{fig:othchan}) shows the results for
just one set of cluster parameters.

The value of $\Omega_{\rm HI}$ at $z \simeq 5.10$ 
($233 \, \rm MHz$) is not known. For the purposes of this paper we use the
value given by  Storrie-Lombardi {\it et al.} (1996)  which is valid
for $z \le 4.7$.  For $\nu=233 \,\, {\rm MHz}$ the number of detectable
clouds in a cluster field is shown in Figure~(\ref{fig:othchan}).

\section{Discussion and Conclusions} 
In this paper we have investigated the possibility of detecting the
redshifted $21\,\,{\rm cm}$ emission from HI clouds at high
redshift. Such observations are not possible with existing radio
telescopes unless we observe the cloud through a cluster gravitational
lens which magnifies the HI flux.  However, this method has the
disadvantage that it will work only for clouds which lie very close to
the caustics of cluster lenses.

Given the lack of a clear picture about the nature of these
objects we have used two simple models, namely the UD and
the ED models (\S 2).  The results are 
similar for both the models (Figures~\ref{fig:ud3} and~\ref{fig:ed3}).       
 Both the models have free parameters which effectively allow us the
freedom of distributing the 
total HI available in the HI clouds at high redshift
into either a large number
of clouds with a small amount of HI each or a small number of clouds with
large amounts of HI each. 
In the former scenario the probability of a cloud being
located very close to  the caustic of a gravitational lens and
experiencing a large magnification is higher but
this is offset by the fact that   the unlensed flux of the HI clouds will be
very  small and  a magnification of $\simeq 50$
may not be sufficient to render the lensed flux above the threshold
value for detection. The parameter range where the probability of a
detection is maximum changes with the threshold flux and this may lead
to  the possibility of using such observations to constrain the
distribution of the HI masses of the clouds. This possibility has not been
studied here in any detail.

We find that for observations at $320 \,\, {\rm MHz}$ with  a detection
threshold of $50 \,\,\mu {\rm Jy}$ centered on a cluster with $\sigma=1200
\,\, {\rm km \, s^{-1}}$,   the chance of detecting a HI cloud is
greater than $10 \%$ provided the minimum HI 
content of these clouds is in the mass 
 range  $5 \times 10^8\,\,{\rm  M}_{\odot} \le  
\M \le 2.5 \times 10^{10} \,\,{\rm M}_{\odot}$. 
The chances of detecting an HI cloud increases for a cluster with a
higher velocity dispersion, and a single deep image in the field of a
cluster such as Abell 1689 (Tyson \etal 1990; Pierre \etal 1991; Smail
\etal 1991), which has $\sigma =1989 \,\, {\rm km \, s^{-1}}$ and
$z_l= 0.196$, might either reveal a cloud or put meaningful bounds on
the mass range and mass spectral index $\alpha$ of the HI clouds.

At $320 \,\, {\rm MHz}$ the GMRT will reach a sensitivity of $50
\,\,\mu {\rm Jy}$ with $100\,\,{\rm hrs}$ of integration at a frequency
resolution of $\Delta \nu = 1.25 \times 10^5 \,\, \rm  Hz$ 
corresponding  to a velocity  width $\Delta V
\simeq 115 \,\, \rm  km \, s^{-1}$.
An alternative strategy would be to observe several clusters to a
threshold flux of $100 \,\, \rm \mu Jy$. The latter strategy may be
superior for a part of the parameter range shown in
Figures~\ref{fig:ud3}, \ref{fig:ed3} and~\ref{fig:othchan}, i.e. the
number of expected detections will be higher for the same total
observation time.

Future radio telescope Square Kilometer Array (SKA) will reach
sensitivity $\simeq 2 \,\, \rm \mu Jy$ at $\nu \simeq 320 \,\, \rm MHz$ for
$\Delta V \simeq 200 \,\, \rm km \, s^{-1}$ in an integration time
of $\simeq 8 \,\, \rm hrs$ \footnote{for details see {\tt
http://www.nfra.nl/skai/science}}. In this case 
observations centered on clusters with $\sigma=1200 \,\,{\rm km \, s^{-1}}$
are expected to detect a few HI clouds  even if
the clouds  have low HI masses in the range  $2 \times 10^7\,\, {\rm M}_{\odot}
\le M_{\rm HI} \le  10^9 \,\,{\rm  M}_{\odot}$.

Our results indicate that the probability of detecting a
gravitationally lensed HI cloud at $233$ and $610 \,\, {\rm MHz}$ is
generally lower than the probability at $320 \,\, {\rm MHz}$. However,
it should be borne in mind that there are larger uncertainties in our
predictions at $233$ and $610 \,\, {\rm MHz}$ as compared to the
predictions for $320 \,\, {\rm MHz}$.
        
We conclude by noting that 
the strategy of carrying out deep radio observations
centered on known cluster lenses will make it possible to detect 
the redshifted 21 cm  emission from lensed  HI clouds with either the
GMRT or the SKA, depending on the HI content of these clouds. Such
observations will shed new light on these objects whose HI content has
been observed only in absorption to  date. 

\acknowledgments
The authors would like to thank Jasjeet Bagla, Jayaram Chengalur,
Divya Oberoi, and  Somak Raychaudhury  for useful discussions and
an anonymous referee for helpful comments.
T.D.S thanks CSIR for providing financial support.

\section*{Appendix A}
Here we present  a derivation of Eq.~(\ref{eq:a1}) which gives 
the specific intensity  of the redshifted HI  emission from a cloud
which has  HI column density $N_{\rm  HI}$, velocity dispersion
$\Delta V$ and is at  a redshift $z$.

The HI contained in a part of the cloud which subtends a solid angle 
 $\Delta \Omega$ at the observer is  $N_{\rm HI} r^2_A(z)
 \Delta\Omega$,  where  $r^2_A(z)$ is   the angular diameter  
distance to the HI cloud. In the rest frame of the cloud the total
luminosity of the HI emission is 
\begin{equation}
\Delta L= A_{21} h_P \nu_e f N_{\rm HI} r^2_A(z) \Delta\Omega\,\,,
\label{eq:app1}
\end{equation}
and it will be  distributed 
over the frequency interval $\Delta \nu_e=\Delta V \nu_e/c$.  
Here $\nu_e=1420 \,\,{\rm MHz}$ is the rest frame frequency of the HI
emission, $h_P$ is the Planck constant,  $f$ is the fraction of HI atoms in
the excited state and  $A_{21}=2.85 \times 10^{-15}\,\, s^{-1}$ is the
Einstein coefficient for the spontaneous emission. 

At the observer the radiation  will be redshifted to a frequency
$\nu=\nu_e/(1+z)$ and it will be distributed over the frequency
interval $\Delta \nu= \Delta \nu_e/(1+z)$ The flux $\Delta F$ 
is given by  
\begin{equation}
\Delta F= \frac{\Delta L}{4 \pi r_L^2(z)}\,\,,
\label{eq:app2}
\end{equation}
where  $r_L(z)$ is the luminosity distance to the HI cloud.
We use this to calculate  the specific intensity 
\be
I_{\nu}=\frac{\Delta F}{\Delta\nu \Delta\Omega} \,\,.
\ee
Using Eqs.~(\ref{eq:app1}) and~(\ref{eq:app2})  and the relation
(see e.g. Peebles 1993)  
\be
\frac{r_A(z)}{r_L(z)}=(1+z)^{-2} \,\, ,
\ee
we obtain 
\be 
I_{\nu}= \frac{N_{\rm HI} \, f \, A_{21} \, h_P c}{4 \pi \, (1+z)^3 \,  
\Delta V  }\,\,.   
\label{eq:app3}
\ee

In a situation where the spin temperature $T_{s}$ is high ($T_{s} \gg
h_P \nu_e/k_B)$ we have  $ f=3/4$, where  $k_B$ denotes the Boltzmann
constant.    Equation (\ref{eq:app3}) can
be evaluated to give
\be
I_{\nu}= 40 \,\, \mu  {\rm Jy \,  arcsec^{-2}} (1+z)^{-3} 
\left( \frac{N_{\rm HI}}{10^{22} \,{\rm cm^{-2}}} \right)  \, 
\left(  \frac{200 \,\, {\rm km \, s^{-1}}}{\Delta V}  \right) \,.
\ee

\newpage

\end{document}